\documentclass[useAMS,usenatbib]{mn2e} 
\usepackage{aas_macros}
\usepackage{graphics}
\usepackage[pdftex]{graphicx}
\usepackage{epstopdf}
\usepackage{epsfig}  
\usepackage{natbib} 
\usepackage{float}
\usepackage{amsmath}
\usepackage{times}
\usepackage[varg]{txfonts}
\usepackage{verbatim} 
\usepackage{multirow,bigdelim} 
\usepackage{color}
\usepackage{vmargin}
\setmarginsrb{1.2cm}{1cm}{1.2cm}{1cm}{1cm}{1cm}{1cm}{1cm}

\def \eg\ {{\it e.g.}\ }

\def \LCDM{$\Lambda$CDM}
\def \msun {M$_{\odot}$}

  \makeatletter
    \renewcommand{\paragraph}{\@startsection{paragraph}{4}{\z@}%
      {-3.25ex\@plus -1ex \@minus -.2ex}%
      {1.5ex \@plus .2ex}%
      {\normalfont\small\centering}}
     
    \renewcommand{\subparagraph}{\@startsection{subparagraph}{5}{\z@}%
      {-3.25ex\@plus -1ex \@minus -.2ex}%
      {1.5ex \@plus .2ex}%
      {\normalfont\small\centering}}
    \makeatother

\newcommand{\eagle}{{\sc eagle}}

\newcommand{\kms}{{ km~s$^{-1}$}}

\title[BTFR \& galaxy sample]{The Baryonic Tully-Fisher Relation cares about the Galaxy Sample}
\author[Sorce \& Guo]
{Jenny G. Sorce\thanks{E-mail: \text{jsorce@aip.de}}, 
Quan Guo\\
Leibniz-Institut f\"{u}r Astrophysik, 14482 Potsdam, Germany\\
}

\begin{document}

\date{}

\pagerange{\pageref{firstpage}--\pageref{lastpage}} \pubyear{2015}

\maketitle

\label{firstpage}

\begin{abstract}
\indent 
The Baryonic Tully-Fisher relation (BTFR) is a clear manifestation of the underlying physics of galaxy formation. As such, it is used to constrain and test galaxy formation and evolution models. Of particular interest, apart from the slope of the relation, is its intrinsic scatter. In this paper, we use the \eagle\ simulation to study the dependence of the BTFR on the size of the simulated galaxy sample. The huge number of datapoint available in the simulation is indeed not available with current observations. Observational studies that computed the BTFR used various (small) size samples with the only obligation to have galaxies spanning over a large range of masses and rotation rates. Accordingly, to compare observational and theoretical results, we build a large number of various size datasets using the same criterion and derive the BTFR for all of them. Unmistakably, their is an effect of the number of galaxies used to derive the relation. The smaller the number, the larger the standard deviation around the average slope and intrinsic scatter of a given size sample of galaxies. This observation allows us to alleviate the tensions between observational measurements and $\Lambda$CDM predictions. Namely,   the size of the observational samples adds up to the complexity in comparing observed and simulated relations to discredit or confirm $\Lambda$CDM. Similarly, samples, even large, that do not reflect the galaxy distribution give on average biased results. Large size samples reproducing the underlying distribution of galaxies constitute a supplementary necessity to compare efficiently observations and simulations.
\end{abstract}


\section{Introduction}

The underlying physics of galaxy formation is visible in the empirical two-or-more galaxy parameter correlations. Among these relations, the Baryonic Tully-Fisher relation \citep[hereafter BTFR, e.g.][]{2000ApJ...533L..99M,2001ApJ...550..212B,2004PASA...21..412G,2005ApJ...632..859M,2005A&A...431..511P,2008MNRAS.386..138B,2009A&A...505..577T,2009AJ....138..392S,2010AJ....140..663G,2014AJ....147..134Z}, extension of the Tully-Fisher relation \citep[hereafter TF, e.g.][]{1977A&A....54..661T,1988ApJ...330..579P,1995ApL&C..31..263T,1997AJ....113...22G,1999ApJ...513..561C,2000ApJ...533..744T,1997AJ....113...53G,2002A&A...396..431K,2006MNRAS.373.1125B,2007MNRAS.381.1463N,2007ApJS..172..599S,2010MNRAS.409.1330W,2012ApJ...749...78T,2012MNRAS.425..296M,2013MNRAS.433.2667R,2013MNRAS.432.3085T,2013ApJ...765...94S,2014MNRAS.444..527S} to the low luminosity galaxies, is often used to constrain and test galaxy formation and evolution models, with various computational methods, in the $\Lambda$CDM scenario as well as in alternative theories \citep[e.g.][]{1999ApJ...513..555S,1998MNRAS.295..319M,1998ApJ...499...41M,1995ApJ...439..520E,2000ApJ...530..177V,2004MNRAS.354..477M,2007ApJ...671.1115G,2007MNRAS.374.1479G,2008AJ....136.1340A,2012AJ....143...40M,2013MNRAS.434.3142A,2012MNRAS.424.3123D,2015MNRAS.454..322D}. While the TF links galaxy luminosities to their rotation rates, the BTFR adds galaxy gas in the equation to correlate galaxy baryonic masses and rotation rates. The baryonic mass (M$_{bar}$) of a galaxy is defined as the sum of its gas (M$_g$) and stellar  (M$_*$) masses. 

Of particular interest are the slope and the intrinsic scatter of the BTFR as they impose strong constraints on galaxy formation and evolution models. A variety of slopes and scatters have been measured over the past years (see references above). The various measured slope and scatter values for the BTFR can mostly be reconciled considering that these measurements are affected by the stellar normalization \citep{2014AJ....147..134Z}, by the rotation rate retained measurements \citep{2012AJ....143...40M} and by the observation uncertainties. The latter, indeed, add up to the measurement (of the galaxy baryonic mass and rotation rate) challenge and to the various methods and calibrations available once the rotation curve is acquired \citep[e.g.][]{1991ApJ...368...60P,1997AJ....114.2402C,1999ApJ...513..561C,2001ApJ...563..694V,2007MNRAS.381.1463N,2009AJ....138.1938C}. While some works attempt to reconcile the different BTFRs by using different method-obtained rotation rates \citep{2008AJ....136.2563W}, some focus on the stellar normalization \citep[e.g.][]{2014AJ....147..134Z} and others disentangle the intrinsic scatter from the total observational scatter of the BTFR \citep[e.g.][]{2007ApJ...671.1115G,2012AJ....143...40M}. While it is clear that these three variables (stellar normalization, rotation-rate measurement and observational uncertainties) affect the slope and the scatter of the BTFR and while works dedicated to deriving the BTFR are always careful to use a wide spanning range of spiral galaxy properties (in terms of both rotation rate and mass) not to bias the relation, there is currently no clear attempt known to us to study the impact of the galaxy sample size on the measured slope and intrinsic scatter of the BTFR.

In this paper, we propose, everything else aside, to investigate the impact of the number of galaxies used to derive the BTFR slope and intrinsic scatter, as well as the effect of the galaxy distribution in the samples. For that purpose, we use the \eagle\ simulation \citep[described briefly hereafter in Section 2]{2015MNRAS.446..521S} as it represents a fair patch of galaxy distribution in the Universe assuming a $\Lambda$CDM cosmological model, ideal for our purposes. In the third section, we study the BTFR using successively all the spiral (defined as such using a selection criterion) galaxies in the simulation and then using reduced size samples and binned samples. In Section 4, we compare our findings with measured BTFR obtained with different datasets in the literature before concluding.


\section{The Hydrodynamical Simulation}

The \eagle\ simulation comprises a set of runs with different box sizes and mass resolutions. Here we use the largest \eagle\ simulation to have a sufficient number of simulated galaxies into our possession to test effectively the impact of the sample size on the BTFR. This simulation follows $1504^3\approx3.4\times10^9$ dark matter particles and the same number of gas
particles in a comoving $100^3$ Mpc$^3$ cubic volume within the framework of a flat \LCDM\ cosmology with parameters taken from the Planck mission \citep{2014A&A...571A..16P}: $\Omega_{\rm m}=0.307$, $\Omega_\Lambda = 0.693$, $\Omega_{\rm b} = 0.04825$, $H_0= 67.77 {\rm km\,s^{-1}\,Mpc^{-1}}$(i.e.
$h=0.6777$), $\sigma_8=0.8288$ and $n_{\rm s}= 0.9611$. It has been performed with \textsc{Gadget-3} \citep{Springel2005} modified by  a pressure-entropy formulation of SPH \citep{Hopkins2012}, new prescriptions for viscosity and thermal diffusion \citep[appendix B in][and Dalla Vecchia in prep.]{sch14} and a different time stepping scheme \citep{Durier2012}. A Friends-of-Friends (FoF) algorithm \citep{Davis1985} allowed to identify the halos, while the \textsc{Subfind} code \citep{Springel2001, Dolag2009} gave the bound structures that these latter encompassed. The galaxies are defined as ensembles of stars within these gravitationally bound sub-haloes. 

With the subgrid parameters for feedback calibrated to match the observed $z\sim 0$ galaxy stellar mass function, \eagle\ simulations are able to reproduce a representative set of observed galaxy properties or relations at low redshift, such as specific star formation rates, Tully-Fisher relation and luminosity functions at $z=0.1$ from the ultraviolet to the near-infrared bands \citep{2015MNRAS.452.2879T}.

Simulated galaxies with $u, g, r, i, z$ magnitude bands (the minimum stellar mass is $10^{8.5}\, \textrm{M}_\odot$) are selected from \eagle\ public released halo and galaxy catalogues \citep{McAlpine2015} \footnote{\texttt{http://www.eaglesim.org/database.html}} at redshift $z=0$.  We define the baryonic mass (M$_{bar}$) of simulated galaxies as the sum of the gas and stellar mass within the 30~kpc aperture (\texttt{Mass\_Gas} and \texttt{Mass\_Star}). \citet{sch14} showed that this choice of   fixed 30~kpc 3D aperture yields the similar stellar mass function as a Petrosian aperture. The luminosities of galaxies with stellar mass $<10^{11}~M_{\odot}$ are not sensitive to the exact choice of the aperture, since the fraction of light in the extended halo is much lower \citep{2015MNRAS.452.2879T}.

As for their rotation rate, we take the maximum value of the circular velocity (\texttt{Vmax} from the \texttt{SubHalo} table), defined as V$_{max}=G(M(<r)/r)^{1/2}$, where $M(<r)$ is the total mass enclosed in a sphere of physical radius $r$. We will come back to this choice in Section 4.

  We make another cut at the $u$ band magnitude $M_u=-16$ to ensure the completeness of the sample according to the luminosity function following criteria of most of the observational studies. We are interested in having a large sample of spiral galaxies spanning over a large range of properties notably in mass and in rotational velocity, a cut at a smaller magnitude in absolute value would include irregular dwarfs in the sample while a larger magnitude cut in absolute value would remove the low mass spiral galaxies\footnote{Note that regardless, our main conclusion is unaffected by this magnitude cut.}.  Actually, to exclude elliptical galaxies from the sample or conversely to extract spiral galaxies, we use the specific star formation rate (sSFR) to separate star-forming galaxies from passive ones. We apply a cut at $\log({\mathrm{sSFR}}) = -2.0$ for galaxies at redshift $z=0$ following \citet{fur14}, who based their choice on a compilation of different observational studies. Galaxies above this cut are considered to be star-forming galaxies and by extension spiral galaxies in the rest of this paper in agreement with observational studies. Indeed, \citet[][]{2007ApJ...671L.113L} showed that galaxies with rotational velocities above 120 \kms\ are spiral galaxies unless their star formation rate is low (thus their are removed by the selection criterion) ; galaxies with intermediate rotational rate (between 50 and 120 \kms) are mostly late-type spirals or alternatively massive irregular that follow the TFR ; galaxies with low rotational velocities (below 50 \kms) are mostly dwarfs galaxies but are in a minority (less than 10) in our sample thanks to the cut in magnitude.


\section{The simulated BTFR}

\subsection{Of the Total Sample}

\begin{figure}
\centering
\hspace{-0.8cm}\includegraphics[width=0.5\textwidth]{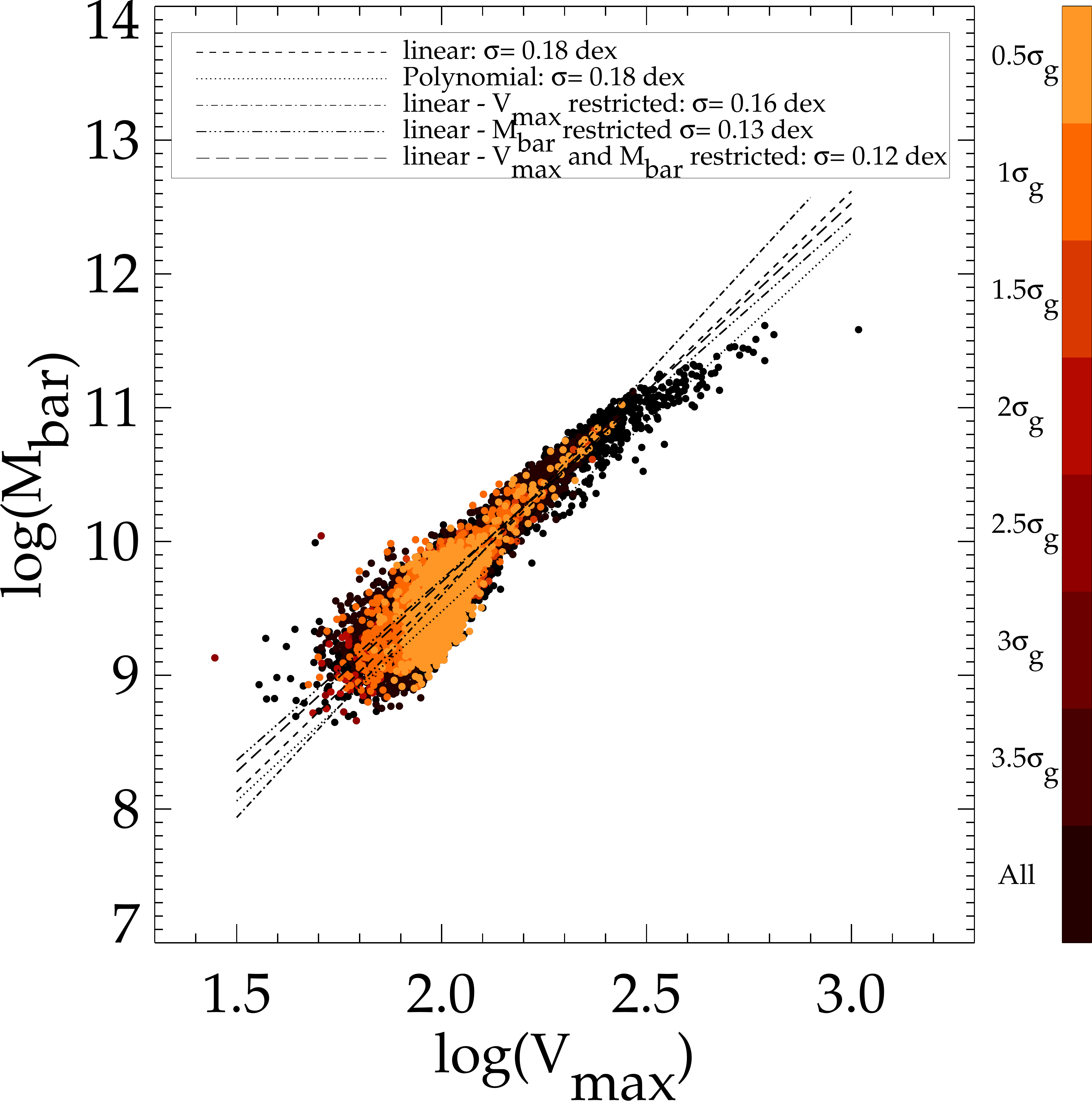}
\caption{Baryonic Tully-Fisher relation for the sample of spiral (selected using log(SFR/M$_*$)$>$-2.0) galaxies in the Eagle simulation (10,225 galaxies) complete down to M$_u$=-16. Dashed and dotted lines are for linear and second order polynomial fits while the dot-dashed, triple-dot-dashed and long-dashed lines fit galaxies with log(V$_{max}$) between 1.9 and 2.5, log(M$_{bar}$) between 9.5 and 11 and these two restrictions together. The legend gives the (intrinsic) scatter $\sigma$ around these different fits. The gradient of color shows the distribution of galaxies assuming that latter to be a 3D-Gaussian with a standard deviation $\sigma_g$.}
\label{fig:TF}
\end{figure}

\begin{figure*}
\centering
\includegraphics[width=0.48 \textwidth]{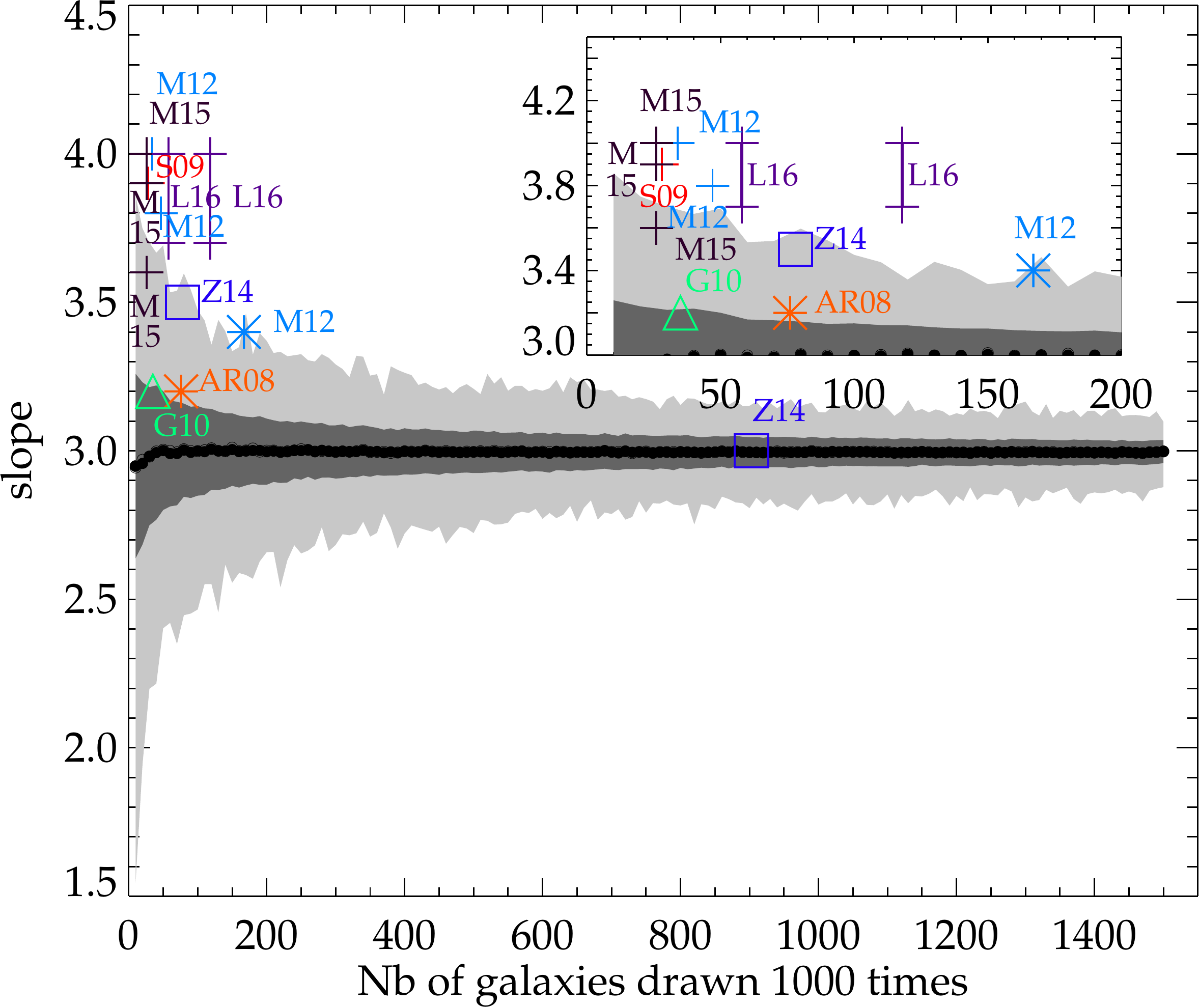}
\includegraphics[width=0.485 \textwidth]{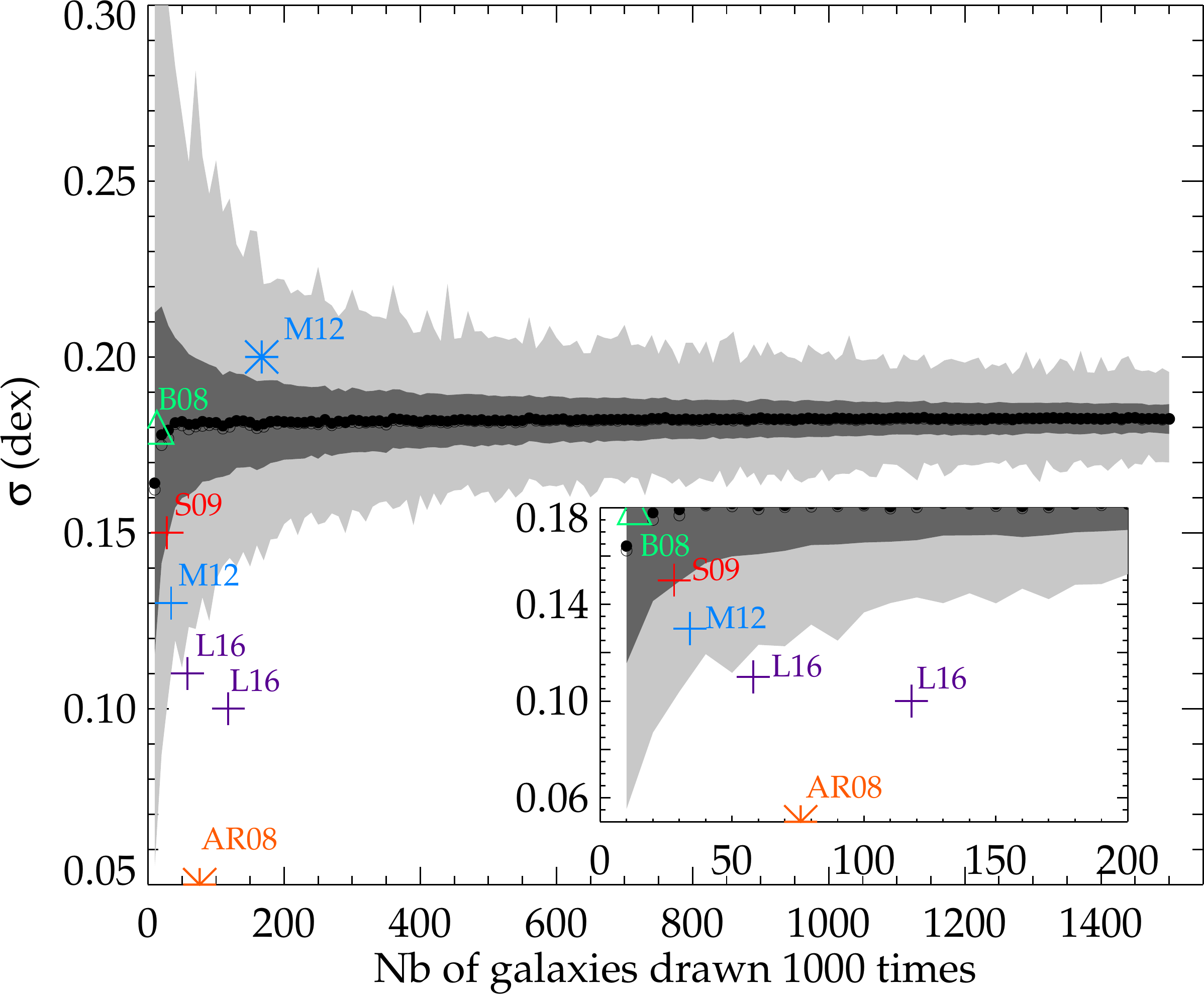}
\caption{Slope (left) and intrinsic scatter (right) of the simulated BTFR fit as a function of the size of the galaxy sample. Filled and open circles show means and medians (when distinguishable from the former). The dark grey area represents the 1$\sigma$ scatter of the slopes and intrinsic scatters for each sample size reproduced 1,000 times. The lighter grey area gives the maximum and minimum for each sample size drawn 1,000 times. Colored symbols (one per type of rotational measurements used) stand for results from observational datasets. The small panels are zooms onto the regions where most of the observational values are.}
\label{fig:sls}
\end{figure*}

\citet{2015MNRAS.446..521S} have already checked the stellar TF relation (correlation between stellar mass M$_*$ and circular velocity $V_{max}$) for the \eagle\ simulation. Here we extend their analysis by introducing the gas mass (M$_g$) in the study and we turn to the BTFR. Using the cut described in section 2., we are `left' with a large sample of 10,225 simulated galaxies that abide by the BTFR plotted on Figure \ref{fig:TF}. One can immediately see that a correlation exists for a large spanning range both in baryonic mass (M$_{bar}$) and in rotational velocity (V$_{max}$). We fit two different relations to the datapoint: a linear fit resulting in a 3.0 dex slope and in a 0.18 dex intrinsic (by definition since simulated data have no observational error) scatter. Following \citet{2014ApJ...792..129N}, we fit also a second order polynomial and recover the same intrinsic scatter value. However, restricting the linear fit to a smaller range of V$_{max}$ (80$<$V$_{max}$$<$315 \kms), the scatter is reduced by 0.02 dex and the slope is increased by 0.3 dex. On the contrary, restricting the fit to a smaller range of M$_{bar}$ (3$\times$10$^{9}$$<$M$_{bar}$$<$10$^{11}$ \kms), the slope is decreased by 0.3 dex while the scatter is still decreased but now by 0.05 dex. Restricting both in velocity and baryonic mass decreases the slope by 0.2 dex and the scatter by 0.06 dex. This observation confirms the importance of using datasets with a wide spanning range of properties to grasp the universal BTFR. We checked that these results are not affected by our `cut for spirals' choice as long as the value of that latter stays reasonable. 

Before comparing these values to those obtained with observations, it is important to note that currently no observational BTFR study is based on a complete sample and above all none reach the number of galaxies used here. This is understandable as gathering accurate rotation rate and baryonic mass measurements constitutes a great observational challenge. For instance, close to face-on galaxies are almost systematically removed from the samples as it is difficult to determine their rotation rate in such conditions. In the next subsection, we look at the simulated BTFR for various size samples of simulated galaxies.

\subsection{Of various-size samples}

While the numerous BTFR observational studies currently available never reached the same number of datapoint as the \eagle\ simulation and while they are never based on complete datasets (a high uncertainty for a galaxy results in its exclusion and there is a limited coverage/overlap of the different required observations), authors are always very careful to use a sample with a large range of velocities and masses to avoid any bias due to a restricted range of parameters and thus to a non-representative sample of the entire population. However, there is no information about the distribution of these galaxies across the different ranges. We then proceed as follows when building our random selection of varying-size datasets from the simulation: we reject a set when galaxies belonging to that latter do not spread over a wide range of velocities and masses. Namely, we impose that there are galaxies with V$_{max}$ $>$ 200 \kms, V$_{max}$ $<$ 60 \kms, M$_{bar}$ $<$ 10$^9$ \msun, M$_{bar}$ $>$ 3$\times$10$^{10}$ \msun\ (note that varying reasonably these values does not affect the results nor does splitting reasonably M$_{bar}$ over M$_*$ and M$_g$). We choose these values in agreement with the histograms of V$_{max}$ and M$_{bar}$ to include the tails of these latter. With this selection criterion in mind, simulated galaxy samples containing from 10 to 1,500 galaxies are drawn from the simulations. We repeat the exercise 1,000 times and we fit a linear BTFR relation to all these galaxy datasets. Figure \ref{fig:sls} gathers the mean (filled circles), median (open circles, when distinguishable from the former), standard deviation (dark grey area), minimum and maximum (of the 1,000 datasets, light grey area) of all the slopes and intrinsic scatters for each given size sample category. 

The result is unmistakable. A too small sample size results in a BTFR slope and an intrinsic scatter that can grandly depend on that very same sample as shown by the large standard deviations around the average BTFR slope and intrinsic scatter. The smaller the sample the larger the probability to measure a different slope from one sample to another. The intrinsic scatter is not spared. Again the standard deviation is very large suggesting that too small samples can give very different results. Table \ref{Tbl:2} actually gathers the values in percentage chance that a measured slope (scatter) may differ from the universal slope (scatter) at 1$\sigma$ and 3$\sigma$ for a given sample size. This table gives a direct access to the uncertainty on a measurement due to the sample size shown on Figure \ref{fig:sls}. We are able to reproduce these findings with the higher resolution \eagle\ simulation for the small sample size using the same criteria, however the total number of galaxies is too low to extend the study also to large samples. Regardless, it is reassuring that the larger standard deviations of the slope and intrinsic scatter for the small size samples are not caused by the resolution of the simulation. Before making some comparisons with observational BTFRs from the literature, we look in the next section at the slope and intrinsic scatter as a function of the circular velocity of the simulated galaxies to understand better how a small sample, although spread over a large range of properties, can give a biased (from the universal relation) result.

\begin{figure*}
\centering
\hspace{-1.3cm}\includegraphics[width=0.38 \textwidth]{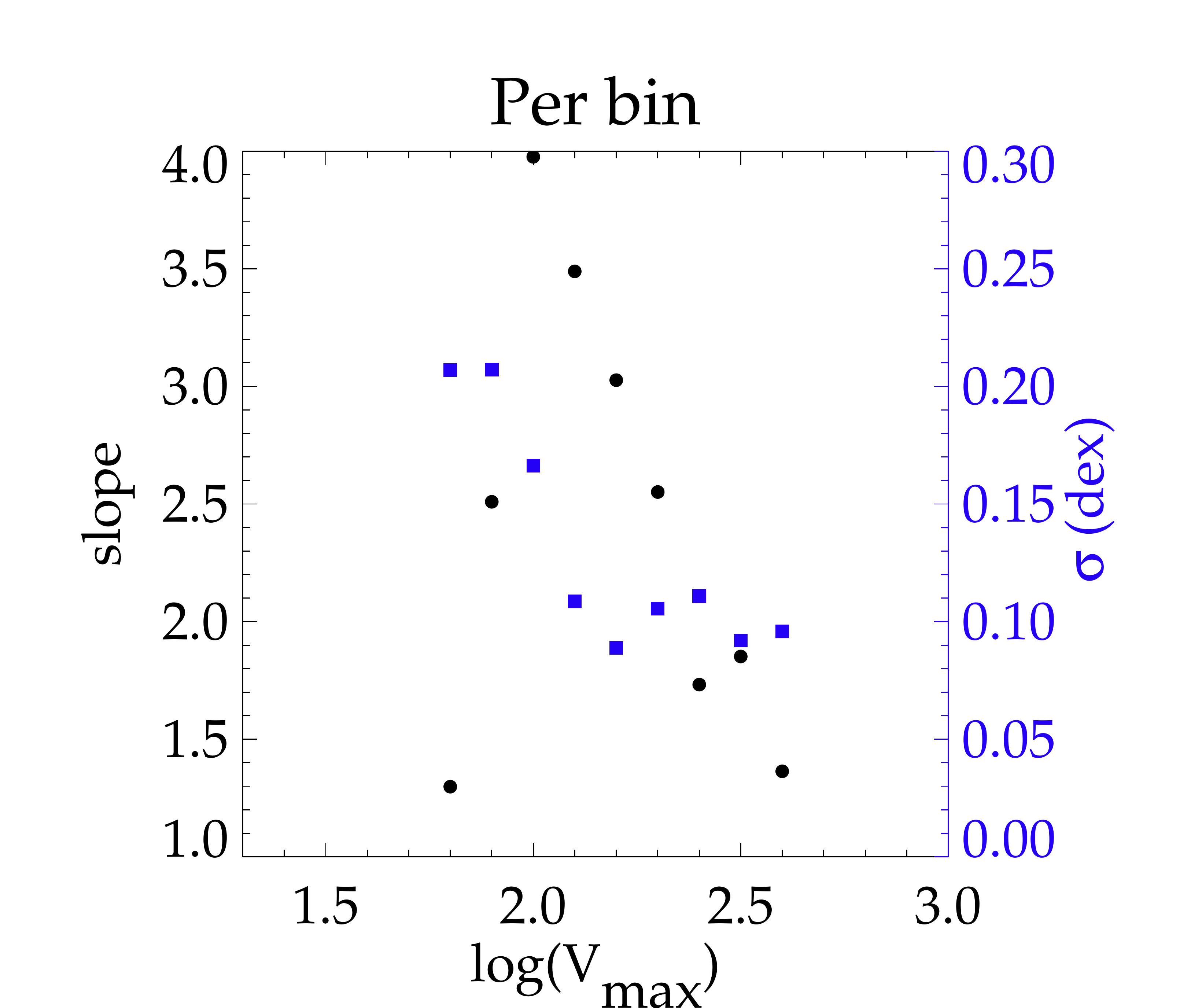}
\hspace{-0.7cm}\includegraphics[width=0.38 \textwidth]{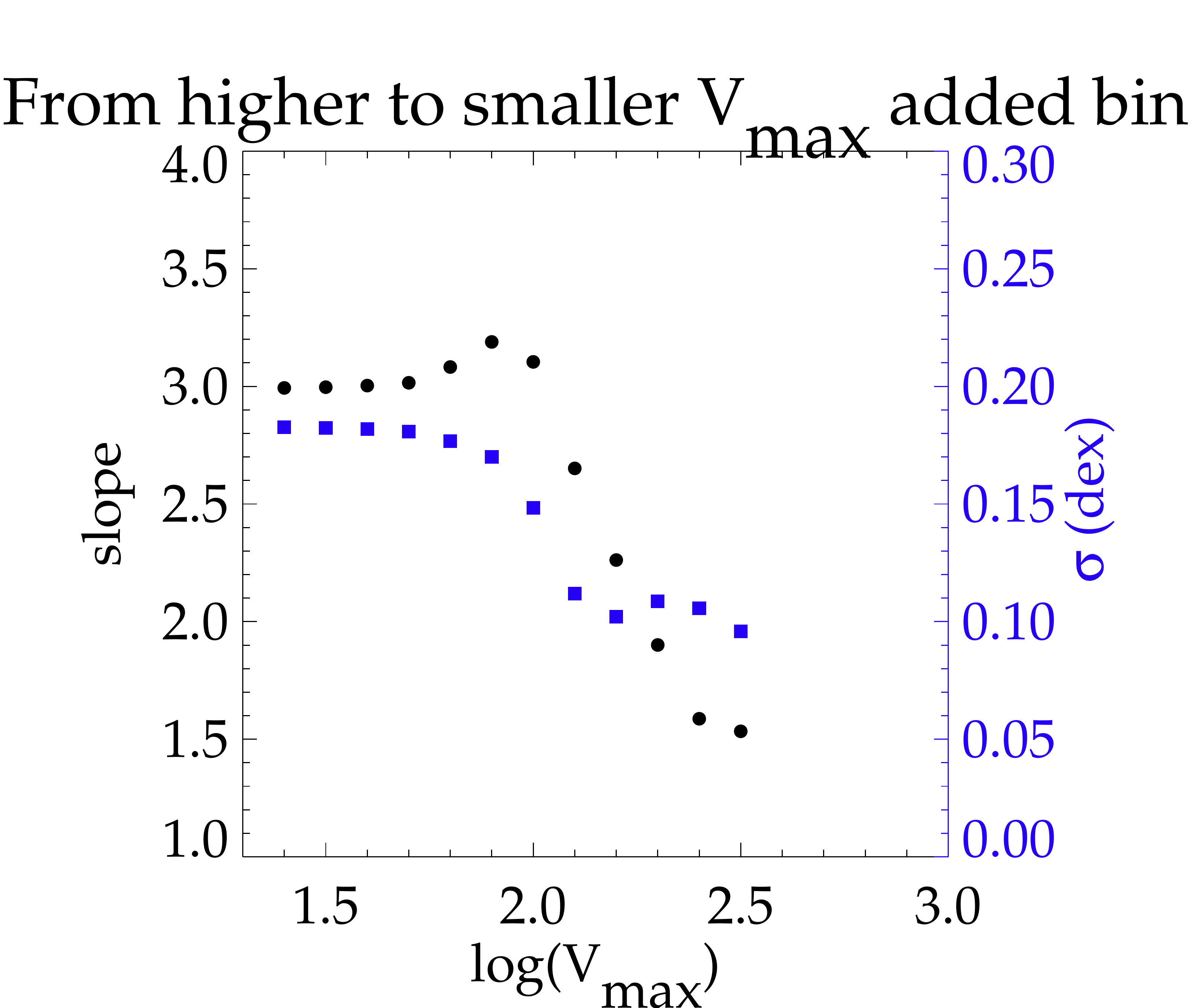}
\hspace{-0.8cm}\includegraphics[width=0.38 \textwidth]{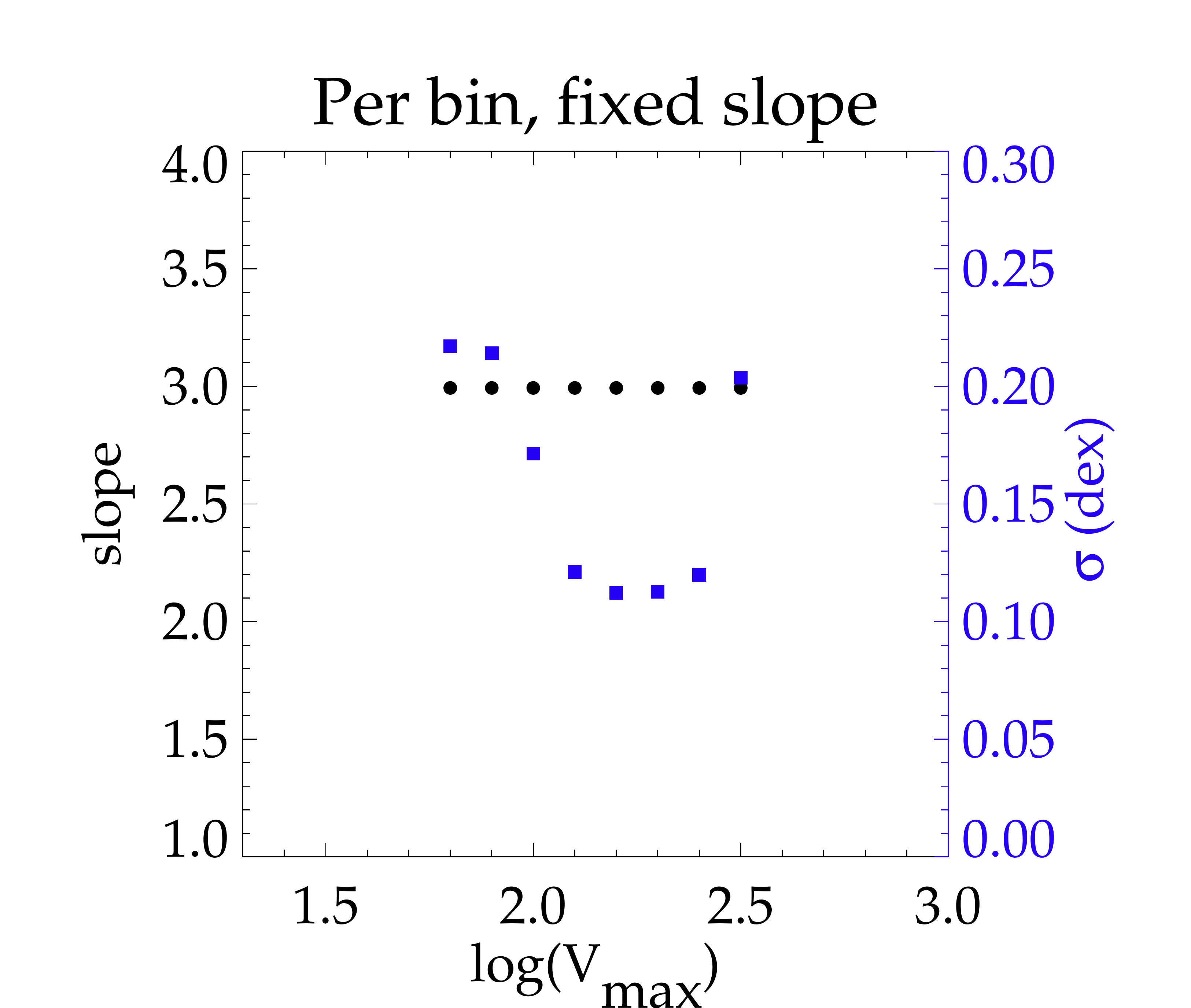}
\caption{Slope (black filled circle) and intrinsic scatter (blue filled square) of the BTFR as a function of the rotation velocity for simulated galaxies per 0.5 dex bin size (left), per cumulative 0.5 dex bin size, adding from the highest bins in terms of velocity to the smallest ones (middle), per 0.5 dex bin size fixing the slope to that of the total sample (right). A value is shown only if at least 6 galaxies contributed to its derivation.}
\label{fig:binned}
\end{figure*}

\subsection{As a function of galaxy parameters}

Observations show that the BTFR has a larger scatter for low mass and velocity end \citep[e.g.][]{2008MNRAS.386..138B}. In this subsection, we then fit the BTFR for galaxies in different bins of velocities to determine whether this claim is reproduced in the simulation. Slopes and intrinsic scatters as a function of the velocity (binsize of 0.5 dex, plotted only if at least 6 galaxies are in the bin) are given in Figure \ref{fig:binned}. The left panel of the figure confirms what was already visible in Figure \ref{fig:TF}, i.e. in agreement with observations, low mass and velocity galaxies are more scattered around the BTFR than their counterparts at higher mass and velocity. The intrinsic scatter (blue filled squares) starts as low as 0.10 dex and increases up to 0.21 dex. Interestingly, the scatter stays quite stable for galaxies with rotation velocities higher than 100 \kms\ ($\sigma$ $\approx$ 0.10 dex) and increases abruptly for galaxies with rotation rates smaller than 100 \kms\ to then reach another plateau. However, processing the galaxies per bin, we are actually restricting the data to small range in velocities and masses, allowing the slopes of the BTFR per bin to vary widely (same panel but black filled circles) from 1.3 to 4 dex. 

In the middle panel of Figure \ref{fig:binned}, we proceed differently by including successively galaxies in the BTFR fit. We start with high rotation velocity galaxies and go down to the low rotation rate ones. The scatter and the slope increase together with the addition of lower and lower rotation rate galaxies. Namely, it confirms that indeed galaxies with a wide spanning range of properties are needed to obtain the slope of the entire sample, however by including more and more low rotation rate galaxies the scatter increases more and more. Again two plateaus are visible with a passage from the lowest one to the highest one at log(V$_{max}$)$\approx$2.0 dex.

Finally, assuming now that we know the slope of the universal BTFR, we look in the right panel of Figure \ref{fig:binned} at the scatter as a function of the rotation velocity with a fixed slope. Again, the scatter is only 0.11-0.12 dex for the high rotation rate galaxies while it rises to 0.2-0.22 dex for the low rotation velocity galaxies. \\

This implies that although a large spanning range of galaxy properties should in principle enable to get the proper universal slope and intrinsic scatter of the BTFR, it could be that a non-equilibrated sample of galaxies (larger or smaller number of galaxies with high rotation rates with respect to those with low rotation rates) might give a somewhat biased result. We study this hypothesis in the next section.

\subsection{As a function of galaxy distributions}
To understand the impact of the galaxy sample distribution on the measured BTFR, we vary the criteria applied to keep or reject a dataset to construct Figure \ref{fig:sls}. Namely, not focusing only on having galaxies spanning over a large range of properties, we observe the following:
\begin{itemize}
\item forcing to have approximately the same number of galaxies per 0.5 bin in log(V$_{max}$), the average slope varies from 2.9 to 2.7 dex from the small to the large samples and the average scatter varies from 0.15 to 0.19 dex. This implies that an homogeneous distribution across the bins is not adequate to recover the universal BTFR. This is understandable since the galaxies are not homogeneously distributed over the different range of properties.
\item forcing to have a higher number of galaxies in the higher bins of log(V$_{max}$), the average slope varies from 3.0 to 2.7 dex and the average scatter from 0.145 to 0.175 dex. 
\item forcing to have a higher number of galaxies in the smaller bins  of log(V$_{max}$), the average slope varies from 2.9 to 2.7 dex and the average scatter from 0.14 to 0.2 dex.
\item forcing to have a distribution of galaxies in the different bins that reproduces the global one, the average slope varies between 3. and 2.95 dex and the scatter from 0.16 to 0.19 dex.
\end{itemize}
\vspace{-0.04cm}

The standard deviations around these average values are still well larger for the smallest size samples than for the largest ones. Still, one can conclude that picking randomly the galaxies ensuring only that galaxy properties span over a large range of values is similar to forcing the distribution of properties in the sample to match that of the entire sample once both are normalized. This is reassuring in the sense that it means selecting randomly the simulated galaxies with the sole constraint of having a wide range of properties, we did not bias the result. However, it is worth noticing that if in a sample there are more high or low velocity galaxies, the sample is biased, even if it is a large sample. It is also important to see that trying to have the same number of galaxies per velocity bins also biases the result. Then the best way to proceed is too pick randomly galaxies provided that this selection is not based on an already biased sample. This is possible for the simulation as all the galaxies are available. This is a priori also possible for observations as galaxies  are randomly selected to be observed, although one has to be careful here since observations always favor the brightest galaxies. Then it could be that although galaxies with various properties are constituting an observational sample, this sample is biased not only because it is small but because it has more galaxies in higher or lower (although the first case is more likely than the first one) velocity bins. 

This implies that not only the galaxy sample size should be large enough but also the galaxies distribution should reproduce the underlying one. In light of this information, we compare different observational results with those obtained when selecting different size samples spanning over a wide range of properties keeping in mind that for the simulation, it also means reflecting the underlying galaxy distribution.

\begin{table*}
\begin{center} 
\begin{tabular}{ccccccccccc}
\hline
\hline
Nb galaxies & V$_{type}$ & slope &  intrinsic scatter & short name &&Nb galaxies & V$_{type}$ & slope &  intrinsic scatter & short name \\
\hline
13 & W$_{20}$&	&0.18$^*$	& B08	&&903   &W$_{50}$&3.0&		        & Z14 \\
76 &  V$_{m}$& 3.2	&0.05	& AR08	&& 78   &W$_{50}$& 3.5&  		& Z14 \\ 
28 & V$_{f}$   & 3.9	&0.15$^*$	& S09	&&26& V$_{f}$	& 3.9		&		& M15	\\
35 &W$_{20}$& 3.2  &   		& G10 	&&26& V$_{f}$	& 4.0		&		& M15 	\\
47 & V$_{f}$   & 3.8  	&		& M12 	&&26& V$_{f}$	& 3.6		&		& M15	\\
34 & V$_{f}$   & 4.0   & 0.13 	& M12	&&58 &V$_{f}$    & 3.7-4.0$^\dag$&0.11	& L16 \\ 
167&W$_{20}$& 3.4	&  0.2        & M12	&&118&V$_{f}$   &  3.7-4.0$^\dag$& 0.10 & L16\\
\hline
\hline
\end{tabular}
\end{center}
\caption{Properties of the observational datasets when available from the literature. (1) and (6)  Number of galaxies in the sample, (2) and (7) type of velocity measurements, either line width W at 20 or 50 \% of the HI-profile peak or maximum, m, and flat, f, velocity V (3)  and (8) slope of the BTFR, in dex, (4) and (9) intrinsic scatter of the BTFR, in dex, (5) and (10) Short names used in Figure \ref{fig:sls}. B08 stands for \citet{2008MNRAS.386..138B}, AR08 for \citet{2008AJ....136.1340A}, S09 for \citet{2009AJ....138..392S}, G10 for \citet{2010AJ....140..663G}, M12 for \citet{2012AJ....143...40M}, Z14 for \citet{2014AJ....147..134Z}, M15 for \citet{2015ApJ...802...18M} and L16 for \citet{2016ApJ...816L..14L}. $^*$ these numbers are from \citet{2012AJ....143...40M}. $^\dag$ range of slope values for stellar normalization coefficients that results in the smaller intrinsic scatter.}
\label{Tbl:1}
\end{table*}

\begin{table*}
\begin{center} 
\begin{tabular}{r@{ }@{ }@{ }r@{ }@{ }@{ }r@{ }@{ }@{ }r@{ }@{ }@{ }r@{ }@{ }@{ }@{ }@{ }@{ }@{ }@{ }@{ }@{ }@{ }r@{ }@{ }@{ }r@{ }@{ }@{ }r@{ }@{ }@{ }r@{ }@{ }@{ }r@{ }@{ }@{ }@{ }@{ }@{ }@{ }@{ }@{ }@{ }@{ }r@{ }@{ }@{ }r@{ }@{ }@{ }r@{ }@{ }@{ }r@{ }@{ }@{ }r@{ }@{ }@{ }@{ }@{ }@{ }@{ }@{ }@{ }@{ }@{ }r@{ }@{ }@{ }r@{ }@{ }@{ }r@{ }@{ }@{ }r@{ }@{ }@{ }r@{ }@{ }@{ }@{ }@{ }@{ }@{ }@{ }@{ }@{ }@{ }r@{ }@{ }@{ }r@{ }@{ }@{ }r@{ }@{ }@{ }r@{ }@{ }@{ }r@{ }@{ }@{ }@{ }@{ }@{ }@{ }@{ }@{ }@{ }@{ }}\hline
\hline
Nb & \multicolumn{2}{l} {$\sigma_{slope}$} & \multicolumn{2}{l} {$\sigma_{scatter}$} &Nb & \multicolumn{2}{l} {$\sigma_{slope}$} & \multicolumn{2}{l} {$\sigma_{scatter}$} &Nb & \multicolumn{2}{l} {$\sigma_{slope}$} & \multicolumn{2}{l} {$\sigma_{scatter}$} &Nb & \multicolumn{2}{l} {$\sigma_{slope}$} & \multicolumn{2}{l} {$\sigma_{scatter}$} &Nb & \multicolumn{2}{l} {$\sigma_{slope}$} & \multicolumn{2}{l} {$\sigma_{scatter}$} \\
&  \multicolumn{2}{l} {in \% at} &  \multicolumn{2}{l} {in \% at} &&  \multicolumn{2}{l} {in \% at} &  \multicolumn{2}{l} {in \% at} &&  \multicolumn{2}{l} {in \% at} &  \multicolumn{2}{l} {in \% at} &&  \multicolumn{2}{l} {in \% at} &  \multicolumn{2}{l} {in \% at} &&  \multicolumn{2}{l} {in \% at} &  \multicolumn{2}{l} {in \% at} \\
& 1$\sigma$ & 3$\sigma$ & 1$\sigma$ & 3$\sigma$ && 1$\sigma$ & 3$\sigma$ & 1$\sigma$ & 3$\sigma$ && 1$\sigma$ & 3$\sigma$ & 1$\sigma$ & 3$\sigma$ && 1$\sigma$ & 3$\sigma$ & 1$\sigma$ & 3$\sigma$ && 1$\sigma$ & 3$\sigma$ & 1$\sigma$ & 3$\sigma$   \\ 
\hline
       10     &     10     &     32     &     30     &     90     &310     &     2     &     8     &     4&    14       &  610     &     2     &     6     &     3     &     10     &  910     &     1     &     5     &     2     &     8     &    1210     &     1     &     4     &     2     &     7     \\     
    20     &     9     &     27     &     18     &     56     & 320     &     2     &     8     &     4 &    13     &    620     &     2     &     6     &     3     &     10     &  920     &     1     &     5     &     2     &     8     &    1220     &     1     &     4     &     2     &     7     \\     
    30     &     8     &     25     &     16     &     48     & 330     &     2     &     8     &     4 &    13     &    630     &     2     &     6     &     3     &     10     &  930     &     1     &     5     &     2     &     8     &    1230     &     1     &     4     &     2     &     7     \\     
    40     &     7     &     22     &     13     &     39     & 340     &     2     &     8     &     4 &    13     &    640     &     2     &     6     &     3     &     10     &  940     &     1     &     4     &     2     &     8     &    1240     &     1     &     4     &     2     &     7     \\     
    50     &     6     &     19     &     11     &     35     & 350     &     2     &     8     &     4 &    14     &    650     &     1     &     5     &     3     &     9     &    950     &     1     &     5     &     2    &     8     &  1250     &     1     &     4     &     2     &     7     \\     
    60     &     6     &     18     &     11     &     33     & 360     &     2     &     8     &     4 &    13     &    660     &     1     &     5     &     3     &     10     &    960     &     1     &     4     &     2    &     8     &  1260     &     1     &     4     &     2     &     7     \\     
    70     &     5     &     17     &     10     &     30     & 370     &     2     &     7     &     4 &    13     &    670     &     1     &     5     &     3     &     9     &  970     &     1     &     5     &     2     &     8     &    1270     &     1     &     4     &     2     &     7     \\     
    80     &     5     &     16     &     9     &     29     &  380     &     2     &     7     &     4  &    13    &    680     &     2     &     6     &     3     &     10     &    980     &     1     &     4     &     2    &     8     &  1280     &     1     &     4     &     2     &     7     \\     
    90     &     5     &     15     &     8     &     26     &  390     &     2     &     7     &     4  &    13    &  690     &     1     &     5     &     3     &     10     &     990     &     1    &     4     &     2    &     8     &    1290     &     1     &     4     &     2     &     7     \\     
  100     &     4     &     14     &     8     &     25     & 400     &     2     &     7     &     4   &    13   &     700     &     1    &     5     &     3     &     10     & 1000     &     1     &     4     &     2     &     8     & 1300     &     1     &     4     &     2     &     7     \\     
  110     &     4     &     14     &     8     &     25     & 410     &     2     &     7     &     4   &    12   &  710     &     1     &     5     &     3     &     9     &  1010     &     1     &     5     &     2     &     8     &    310     &  1     &     4     &     2     &     7     \\     
  120     &     4     &     13     &     7     &     23     & 420     &     2     &     7     &     4   &    12   &  720     &     1     &     5     &     3     &     9     &  1020     &     1     &     4     &     2     &     8     & 320     &     1     &     4     &     2     &     7     \\     
  130     &     4     &     13     &     7     &     23     & 430     &     2     &     7     &     4   &    12   &  730     &     2     &     6     &     3     &     9     &  1030     &     1     &     4     &     2     &     8     & 330     &     1     &     4     &     2     &     7     \\     
  140     &     4     &     12     &     6     &     20     & 440     &     2     &     7     &     4   &    12   &  740     &     1     &     5     &     3     &     9     &  1040     &     1     &     4     &     2     &     8     & 340     &     1     &     4     &     2     &     7     \\     
  150     &     4     &     12     &     6     &     20     & 450     &     2     &     7     &     4   &    12   &  750     &     1     &     5     &     3     &     9     &  1050     &     1     &     4     &     2     &     8     & 350     &     1     &     4     &     2     &     7     \\     
  160     &     3     &     11     &     6     &     20     & 460     &     2     &     7     &     4   &    12   &  760     &     1     &     5     &     3     &     9     &  1060     &     1     &     4     &     2     &     7     & 1360     &     1    &     4     &     2     &     6     \\     
  170     &     3     &     11     &     6     &     20     & 470     &     2     &     7     &     4   &    12   &  770     &     1     &     5     &     3     &     9     &  1070     &     1     &     4     &     2     &     7     & 1370     &     1    &     4     &     2     &     7     \\     
  180     &     3     &     11     &     6     &     18     & 480     &     2     &     7     &     3   &    11    &  780     &     1     &     5     &     3     &     9     &  1080     &     1     &     4     &     2     &     7     &    1380     &     1    &     4     &     2     &     6     \\     
  190     &     3     &     11     &     6     &     19     & 490     &     2     &     6     &     4   &    12   &    790     &     1     &     5     &     3    &     9     &  1090     &     1     &     4     &     2     &     8     &     1390     &  1    &     4     &     2     &     7     \\     
  200     &     3     &     11     &     6     &     18     & 500     &     2     &     6     &     3   &    11    &  800     &     1     &     5     &     3     &     9     &  1100     &     1     &     4     &     2     &     7     &1400     &     1     &     4     &     2     &     6     \\     
  210     &     3     &     10     &     5     &     16     &    510     &     2    &     6     &     3 &    11      & 810     &     1     &     5     &     3    &     9     &  1110     &     1     &     4     &     2     &     7     &  410     &     1    &     4     &     2     &     7     \\     
  220     &     3     &     10     &     5     &     17     &    520     &     2    &     6     &     3 &    11      &  820     &     1    &     5     &     3    &     9     &  1120     &     1    &     4     &     2     &     7     &     1420     &    1     &     4     &     2    &     6     \\     
  230     &     3     &     10     &     5     &     17     &    530     &     2    &     6     &     3 &    11      &  830     &     1    &     5     &     3    &     9     &  1130     &     1    &     4     &     2     &     7     &     1430     &    1     &     4     &     2    &     7     \\     
  240     &     3     &     10     &     5     &     16     &    540     &     2    &     6     &     3 &    11      &  840     &     1    &     5     &     3    &     9     &  1140     &     1    &     4     &     2     &     7     &     1440     &    1     &     4     &     2    &     6     \\     
  250     &     3     &     9     &     5     &     16     &  550     &     2     &     6     &     3    &    11   &    850     &     1     &     5     &     2    &     8     &  1150     &     1    &     4     &     2     &     7     &     1450     &    1     &     4     &     2    &     6     \\     
  260     &     3     &     9     &     5     &     15     & 560     &     2     &     6     &     3     &    11  &     860     &     1     &     5     &     2    &     8     & 1160     &     1     &     4     &     2     &     7     &     1460     &    1     &     4     &     2    &     6     \\     
  270     &     3     &     9     &     5     &     15     &  570     &     2     &     6     &     3    &    11   &    870     &     1     &     5     &     2    &     8     & 1170     &     1     &     4     &     2     &     7     &     1470     &    1     &     4     &     2    &     7     \\     
  280     &     3     &     9     &     5     &     15     &  580     &     2     &     6     &     3    &    10  &     880     &     1    &     5     &     2    &     8     &  1180     &     1    &     4     &     2     &     7     &     1480     &    1     &     4     &     2    &     6     \\     
  290     &     3     &     9     &     5     &     15     & 590     &     2     &     6     &     3     &    10 &     890     &     1     &     4     &     2    &     8     & 1190     &     1     &     4     &     2     &     7     &     1490     &    1     &     4     &     2    &     6     \\    
  300     &     2     &     8     &     4     &     14     &  600     &     2     &     6     &     3    &    10  &     900     &     1    &     5     &     2    &     8     &  1200     &     1     &     4     &     2     &     7     &     1500     &    1     &     4     &     2    &     6     \\     
  \hline
\hline
\end{tabular}
\end{center}
\caption{Values in percentage chance that a measured slope (scatter) may differ from the universal slope (scatter) at 1$\sigma$ and 3$\sigma$ for a given sample size. (1), (6), (11), (16) and (21): sample size, number of galaxies ; (2), (7), (12), (17) and (22) 1$\sigma$ uncertainty on the slope due to the sample size, in \% ; (3), (8), (13), (18) and (23) 3$\sigma$ uncertainty on the slope due to the sample size, in \% ; (4), (9), (14), (19) and (24) 1$\sigma$ uncertainty on the scatter due to the sample size, in \% ; (5), (10), (15), (20) and (25) 3$\sigma$ uncertainty on the scatter due to the sample size, in \%}
\label{Tbl:2}
\end{table*}

\section{The observed BTFR}

Table \ref{Tbl:1} gathers several slopes and intrinsic scatters derived for different observational datasets in the literature. It is important to note that several of these datasets overlap with each others so that we are not expecting too different results between the overlapping sets. Namely, these samples share some identical galaxies and map similar local regions of the sky thus we anticipate that the BTFR measured parameters will lie in comparable areas when overplot onto the simulated results as there are not independently obtained. The observation-derived values are actually plotted in Figure \ref{fig:sls} considering the size of the galaxy sample. One can immediately see that unfortunately most of the observational datasets have very often a too small size to be completely trustworthy for comparisons with simulations. They lie in the region where slopes and scatters can vary widely (even with a sample spanning over a wide range of velocity and mass and a galaxy distribution reflecting the underlying one). The sample with the largest number of observed galaxies has on the other hand a slope in perfect agreement with that obtained with simulated galaxies. The intrinsic scatter is unfortunately not directly available but the observational one is 0.29 dex (Zaritsky, private comm.) leaving the possibility of the scatter due to uncertainties to be as high as $\sim$0.23 dex (assuming the intrinsic scatter to be 0.18 dex like in the simulation). These values are perfectly reasonable. \\

On the other hand, observational slopes all (but that obtained with the largest sample) stand on the high end of the values that can be obtained with the simulated galaxy samples. Several issues are probably in cause: 1) the interdependence of the observed datasets, 2) the cosmic variance of a small local patch of the Universe, 3) the different stellar normalizations, 4) the type of measurements for the rotational velocities, 5) the simulation, based on a model that uses our current (limited) knowledge regarding galaxy formation and evolution.

The matter of the overlapping observational samples has to be addressed directly by observers while that of improving hydrodynamical models is more the responsibility of theorists and developers. The cosmic variance between different small patches of the Universe on the other hand will be reduced in the near future thanks to a combination between hydrodynamical models and constrained simulations of the Local Universe. Indeed, constrained simulations unlike typical cosmological simulations stem from a set of observational constraints thus they resemble the local Large Scale Structure \citep[e.g.][]{2009LNP...665..565H,2010arXiv1005.2687G,2014MNRAS.437.3586S,2016MNRAS.455.2078S}. It is thus expected that combined together with hydrodynamical models, regions similar to the observed one will be reproduced to push the comparisons even further. The slope variation engendered by modifications of the stellar normalization has been discussed to a large extent recently by \citet{2016ApJ...816L..14L}. In their paper, the authors show that the slope of the observationally measured BTFR varies between 3.0 to 4.0 dex depending on the stellar normalization coefficient. They settle for coefficients giving slopes in the range 3.7-4.0 that are compatible with a minimum intrinsic scatter of 0.10-0.11 dex but smaller slopes are not excluded.

Finally, there is the delicate question of the rotation rate measurements that differ between observational samples. These distinct measures result unconditionally in various slopes. In Figure \ref{fig:sls}, diverse symbols are actually used to distinguish the datasets as a function of the velocity measurements retained to derive the BTFR slope. Flat velocities (crosses) always give larger slopes than linewidth-derived velocities either at 20\% (stars) or 50\% (squares) of the HI-profile peak or than maximal velocities (triangles). Simulations are not exempt of this issue \citep[e.g.][]{2016arXiv160101026O}. On top of that, measurements between simulations and observations differ since the dark matter halos are directly available only in the former. Some estimates actually tend to show that the factor between simulated and observed velocity values varies with the mass \citep{2012MNRAS.425.2610R,2014MNRAS.439.1897K}. Such studies show that simulated velocities should be reduced more on the high end mass than on the low end one to fit observed values. Such a behavior tends to increase slightly the slopes obtained with the simulated samples reducing some more the discrepancy with the observations. \\

Regardless, overall, observed datasets currently at hands cannot rule out the $\Lambda$CDM scenario used to run the simulation. The existing tensions are clearly alleviated when considering, in addition to the already existing concerns mentioned above, also the sample sizes.\\

\begin{figure}
\centering
\includegraphics[width=0.48 \textwidth]{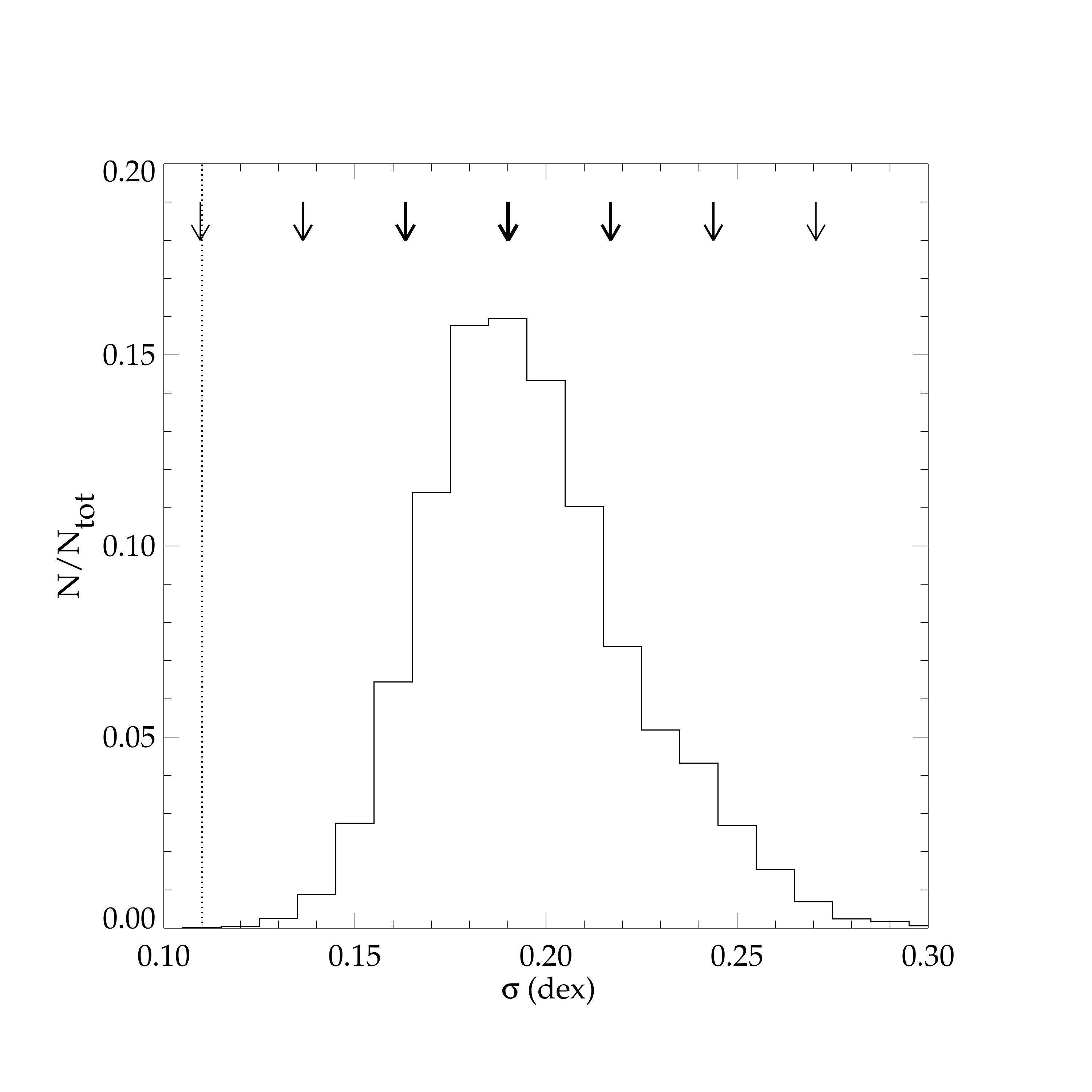}\\
\vspace{-1.5cm}
\includegraphics[width=0.48 \textwidth]{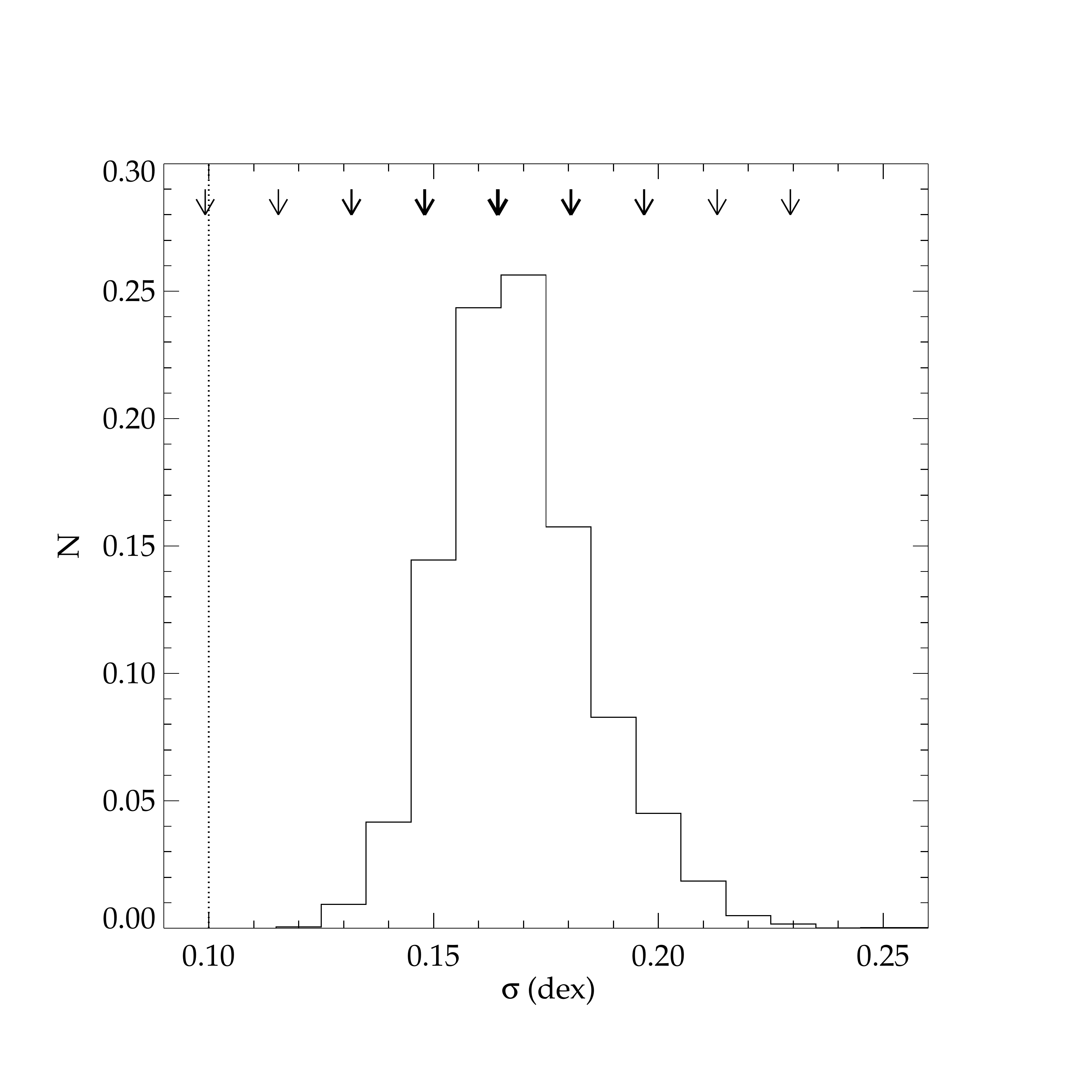}\\
\vspace{-0.8cm}
\caption{Distributions of the intrinsic scatter found for 10,000 simulated galaxy samples reproducing the property distribution of L16 for the small (top) and large (bottom) samples. Arrows from the thickest to the thinnest show the mean and the 1,2 and 3, etc $\sigma$. The dotted lines stand for the observationally measured values.}
\label{fig:L16}
\end{figure}

\begin{figure}
\centering
\includegraphics[width=0.48 \textwidth]{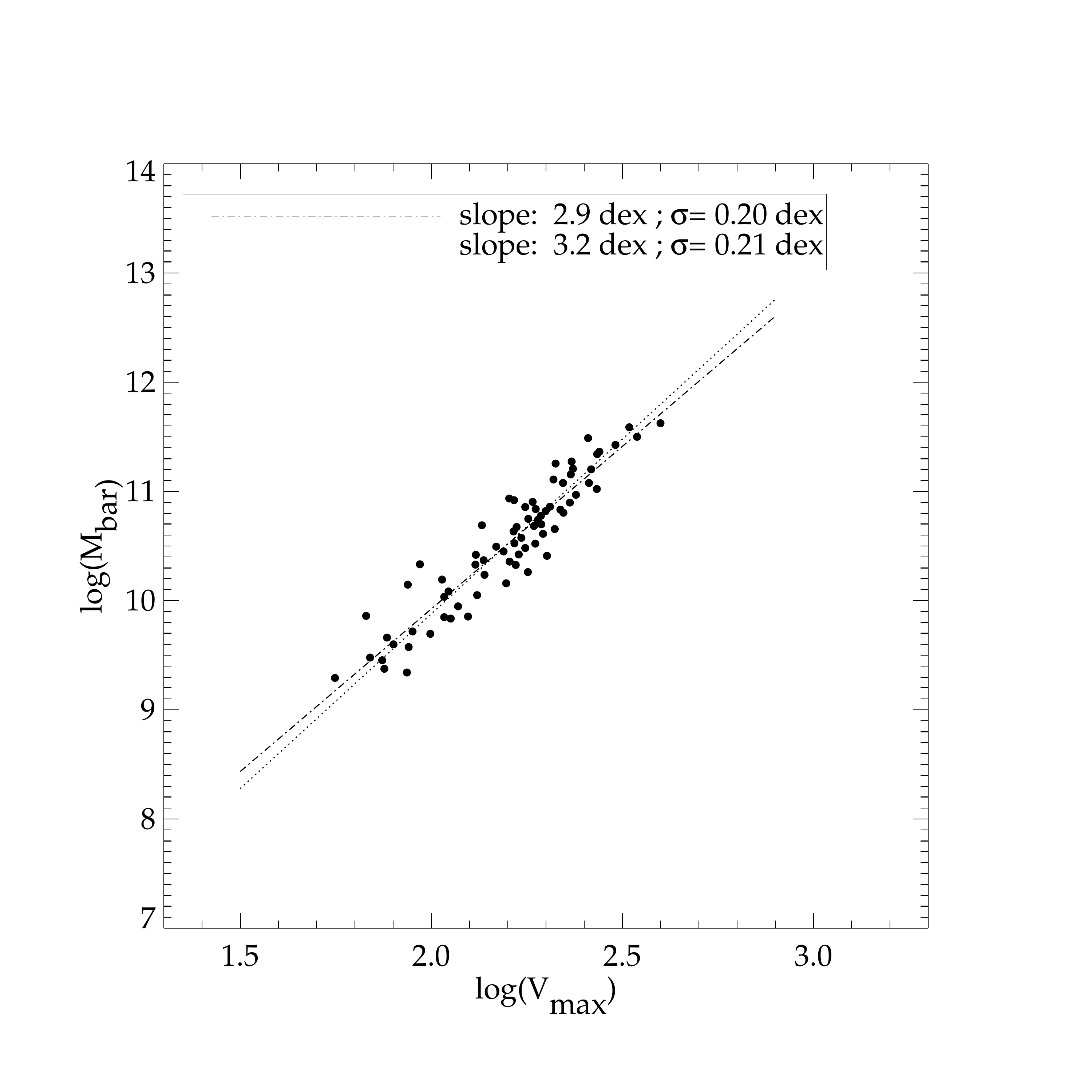}\\
\vspace{-0.8cm}
\caption{Baryonic Tully-Fisher relation for the dataset from AR08. The dotted dashed line gives the linear fit with all the parameters free, while the dotted line shows the fit with a fixed slope.}
\label{fig:AR08}
\end{figure}

Overall, the same ascertainments are valid for the intrinsic scatter. Still, next, we focus some more our attention onto the intrinsic scatter as this is more problematic for the $\Lambda$CDM model than the slope that depends more dramatically on the stellar normalization. \citet{2016ApJ...816L..14L} showed indeed that the intrinsic scatter is also affected by the latter but it is relatively stable and increases only weakly for small normalization coefficients. Interestingly the smaller slope values are also obtained for the smaller coefficients. Typically, to a slope of 3.0 dex corresponds an intrinsic scatter about 0.16-0.17 dex comparable to the values obtained with the simulated samples.

We decide to study in more details the three observational samples that give intrinsic scatters further than 3$\sigma$ away from that obtained with the same number of simulated galaxies. These are those of L16 \citep{2016ApJ...816L..14L} and AR08 \citep{2008AJ....136.1340A}. Even if we have already listed a list of five reasons plus the galaxy sample size has a cause for such an observation, that should really be considered carefully in order to make efficient comparisons in the near future, it is interesting to understand if there exists another culprit, especially since these points are quite below the theoretical predictions. 

In Section 3, a hint at the importance of the galaxy distribution across the range of properties has been highlighted. To study in more details the effect, we bin the three above mentioned observational samples and build three sets of 10,000 simulated galaxy samples, each set sharing one of the three property distributions. For the smaller sample of L16, we find an intrinsic scatter of 0.19$\pm$0.03 dex (minimum of the 10,000 samples 0.11 dex and maximum 0.31 dex) as shown on Figure \ref{fig:L16}, top panel. Proceeding similarly for the larger sample of L16, we get an intrinsic scatter of 0.165 $\pm$ 0.016 dex (minimum of the 10,000 samples 0.11 dex and maximum 0.25 dex), bottom panel of the same Figure. For that of AR08, we measure 0.15 $\pm$ 0.02 dex (minimum 0.10 dex and maximum 0.24 dex). This clearly shows that a larger sample is not necessarily better than a smaller sample in recovering the universal (for the whole simulated galaxy population) intrinsic scatter: in the three examples, the smallest sample give values on average closer to that obtained for the whole sample although with a larger scatter than for the other two larger samples.  AR08-like and large L16-like samples result in low intrinsic scatter on average. Actually, plotting the histograms of properties of AR08 and large L16, we uncover that there are biased towards larger velocity and mass galaxies (i.e. they have more high mass galaxies than low mass ones which is in disagreement with the observed luminosity function), even if they span over a large range of properties. This shows again the importance of not only spanning over a large range of properties but also of reproducing the distribution of the entire population. Regardless, when looking at the AR08 sample, we had some doubts on whether we derive the scatter the same way they did. We believe they proceeded differently. Indeed, taking their datasets by copying datapoint directly from their plot, we obtain in Figure \ref{fig:AR08} a scatter about 0.20 dex and forcing the slope to be theirs, the scatter increases by 0.01 dex confirming our suspicion of different approaches. On the opposite, we are able to reproduce at 0.01 dex the results from L16 implying that we are in agreement with their way of processing the data. Without knowing in details the way they proceed we simply drop AR08's intrinsic scatter value, while the tensions between the L16 values and the simulated values are alleviated some more considering also the distribution of galaxies across the range of properties.


\section{Conclusion}

The correlation between galaxy baryonic mass and rotational velocity, or Baryonic Tully-Fisher relation (BTFR), is a clear proof of the underlying physics of galaxy formation. As such, it is often used to constrain and test galaxy formation and evolution models. Of particular interest apart from the slope of the relation is its intrinsic scatter. In this paper, we have studied the BTFR in the \eagle\ simulation. We measure a slope of 3.0 dex and an intrinsic scatter of 0.18 dex for the entire population of spiral (selected according to a criterion based on the star formation rate and stellar mass, we find that the result is stable for different cuts provided that the chosen value of that latter is reasonable) galaxies. Following observational works that, although they never currently have complete and large galaxy samples (mostly because of observational uncertainties and lack of observations), affirm that it is essential to have samples with a wide spanning range in velocity and mass, we build several size datasets using this sole criterion. There is unmistakably an effect of the sample size on the measured BTFR. The smaller the number of galaxies used to derive the BTFR, the larger the standard deviation around the average slope and intrinsic scatter obtained for a given size sample. A sufficiently large sample (about at least 200 galaxies but the larger the better) of galaxies is requested to obtain a trustworthy, in the sense close to the universal, BTFR. 

Looking for a reason to that observation, we bin the entire population of simulated galaxies in circular velocity and we notice that like in observations low mass and velocity galaxies present a higher scatter around the BTFR. Actually, repeating the same work as before, i.e. building different size datasets, but forcing not only galaxy to span over a wide range of properties but also to be evenly distributed across bins, or to be over-represented in the higher velocity bins or on the contrary into the smaller velocity bins, we discover that even large samples are biased and do not allow to recover the universal (in the sense from the entire simulated population) BTFR relation. However, forcing the sample to reproduce the overall underlying distribution of galaxy properties, we are able to find back the same result as when picking galaxies randomly requesting only from them to cover a large range of properties, i.e. a non biased result for large samples of galaxies. 

Comparing these findings with observation-measured slope and intrinsic scatter values, we immediately notice that a concern arises regarding the dataset sizes. This problem adds up to the already existing list including 1) the interdependence of the observed datasets, 2) the cosmic variance of a small local patch of the Universe, 3) the different stellar normalizations, 4) the type of measurements for the rotational velocities, 5) the simulation, based on a model that uses our current (limited) knowledge regarding galaxy formation and evolution. 

Regardless of the fact that observational slopes all (but that obtained with the largest sample) stand on the high end of the values that can be obtained with the simulated galaxy sample probably because of the long list of issues mentioned above, it is clear that the sample size should also be considered to compare efficiently observed and predicted BTFR values. Currently, datasets at hands are small enough that existing tensions between observations and the $\Lambda$CDM cosmological model are alleviated. Focusing on the intrinsic scatter, as it is a powerful constraint for the $\Lambda$CDM model, we uncover that all but three observation-derived values are well below the simulation-obtained values (i.e. not within 3$\sigma$). Following the hint that results are biased even for large sample if the galaxy distribution does not reflect the underlying one,  we decide to look in more details at these three samples of galaxies. We produce datasets of the same size, a similar spanning range of characteristics and distribution across the property ranges. We show that indeed distribution property-like simulated datasets of the same size results on average in smaller intrinsic scatters than those expected for non distribution property-like, but still spanning over a large range of characteristics, samples of the same size. \\

In summary, the slope and intrinsic scatter of the BTFR depends on a multitude of parameters that all need to be controlled to test efficiently the $\Lambda$CDM cosmological model. To the list that already includes for instance the rotation rate measurement, the stellar normalization and the independence of the datasets, this paper permits to add the size of the sample and the distribution of galaxies that should not only span over a large range of properties but should also reflect the underlying distribution of galaxies in the Universe. We are then able to alleviate the existing tensions between observation-measured values and the $\Lambda$CDM cosmology. We advocate that large size samples not only spanning over a wide range of properties but also reproducing the overall true distribution of galaxy properties are necessary to envisage pushing further the comparisons with simulations. This highlights the importance of dedicating more observing time in producing larger datasets to study the BTFR.

\section*{Acknowledgements}  
JS acknowledges support from the Alexander von Humboldt Foundation. We thank the referee, Stacy McGaugh, for his comments that helped
improved substantially the manuscript. We acknowledge the Virgo Consortium for making their simulation data available.
The \eagle\ simulations were performed using the DiRAC-2 facility at Durham, managed by the ICC, and the PRACE facility Curie based in France at TGCC, CEA, Bruy\`eres-le-Ch\^atel. We thank Dennis Zaritsky for providing us with BTFR values.


\bibliographystyle{mnras}

\bibliography{biblicomplete}

\end{document}